# CONTEST DYNAMICS
# GENERAL BIOMECHANICAL THEORY
# OF THE CONTEST SPORTS

Attilio Sacripanti

University of Tor Vergata Rome  /ENEA Italy/ FILPJK – Italy

## I INTRODUCTION

Contest dynamics as mathematical theory, therefore applicable to all contest sports, is the main topic of this paper. After physical definition of "Athlete" and "Couple of Athlete" systems and after singling out the interaction basic parameter, we analyse the classes of possible potentials describing the interaction and at the end we specify the physical bases of mutual interaction between athletes and the trajectories of flight motion.

All the matter will be connected to measurable quantities or parameters useful for researchers and trainers.

## II "ATHLETE" AND "COUPLE OF ATHLETES" SYSTEMS: DEFINITION AND PHYSICAL CHARACTERIZATION

The physical characterization of the environment of contest easily leads to the individualization of working forces on Athletes systems:

1) gravity force, 2) the push/pull forces 3) constrain reactions of the mat, transferred by friction.

If we define the "Athlete" subsystem as "biomechanical athlete", namely a geometric variable solid of cylindrical symmetry, which takes different positions and performs only definite rotations by the articular joints, then we can easily gave the definition of the global system which concerns the competition analysis: the "Couple of Athletes" system will be defined as a jointed system, of cylindrical symmetry built by the semi-rigid junction of two biomechanical athletes. This system will have only two "energy levels" with specific degrees of freedom.

### A) Couple of Athletes closed system

The two biomechanical athletes have fixed and semi-flexible contact points " the grips".

In this way the two athletes are blended in only one system in stable equilibrium ; this system moves itself by the third principle of dynamics. The ground reaction forces will be, in this case, the resultant of overall push/pull forces produced by both athletes.

***B) "Couple of Athletes" open system.***

The two biomechanical athletes have no fixed contact point , and to keep their condition of unstable equilibrium, they will be at best like a simple inverted pendulum model ( Pedotti 1980 ) (4) while friction will made motion possible by the third principle of Dynamics.

Having defined the Couple of Athletes open and closed systems and its components i.e. the Athletes, the biomechanical analysis of contest dynamics will be studied in terms of system motion and mutual interaction, which according to the principle of the overlap of effects wd be able to be seen separately in order to obtain an easier and more understandable solution .

The mechanics of competition ( not repeatable "situations" which happen "randomly" with small probability of repetition on a very large number of contests) cannot be analysed with the deterministic tools of Newton's mechanics. In effect it would be more useful to study the problem according to statistical mechanics to obtain verifiable experimental results.

**III MUTUAL DISTANCE AS MAIN PARAMETER OF CONTEST DYNAMICS**

The study of "Couple of Athletes" open system easily shows us the main parameter which allows us to classify usefully the mutual position of bodies.

The relative distance between the two Athletes, the attack strategy and the execution of techniques are directly dependent on this parameter.

It is useful to classify three kids of distances which need three different biomechanical approaches.

1) Long distance ( Karate).

It is the distance from which the unarmed Athlete will effect a successful kick attack.

It is the main distance in Karate contests

2) Average distance ( Boxing ).

It is the distance from which it is possible to box

3) Short distance ( Wrestling ).

It is the distance from which it is possible to grip or grasp the adversary. In this condition the Athlete changes his position from unstable to stable . Grips are the main tools to transfer the energy to the adversary both in opposition and in helping the throwing techniques.

# IV REFERENCE SYSTEMS AND INTERACTION : DEFINITION AND CLASSIFICATION

After defining the physical system and the main interaction parameter and specifying the boundary conditions connected to system dynamics, the next step is to define " the reference systems" in which to describe motion and interaction.

Obviously the first reference system will be put in the gym ( a Cartesian reference system solid with the gymnasium walls ) it is in a good approssimation: the inertial reference system or the laboratory reference system.

The second reference system, useful for the simplified study of mutual interaction will be put in the movable barycentre of "Couple of Athletes" open or closed system; this reference system will be called, " Centre of mass reference system".

In all contest sports: interaction can be seen, in the function of mutual distance, as a continuous shortening and lengthening of this parameter, during contest time; plus a few physical specific mechanisms to win for each sports.

These mechanisms can be classified, for contest sports, in two categories..

A) Couple of Athletes closed system

Winning mechanisms able to throw the adversary by two physical principles.

1) application of a couple of forces; 2) application of a physical lever ( Wrestling, Free stile, Greco -Roman stile, Canary stile, Lion stile, Judo, Sambo, Sumo, Koresh, ... ).

So interaction happens by finding a contact point. In the case of the second mechanism , the stopping point, the use of unbalance is necessary.

B) Couple of Athletes open system

Winning mechanisms able to defeat the adversary following the impact theory by direct strokes to the conventional or sensitive points of the body (Boxing, Karate, Kick boxing, Tae Kwon do, Kung fu, Savate, Kendo, Fencing, ... )

So interaction happens by shortening the distance since it becomes useful to stroke the adversary.

# V POSSIBLE CLASSES OF POTENTIALS : A GENERAL STUDY

In each contest sports, interaction is founded on two separate phases, a common one (shortening of mutual distance ) and a specific one ( application of permitted ways to seek advantage: strokes or throwing mechanisms ) The common part is comparable to a classic "two body problem in central field ".From mechanics we remember:

a) instead of studying the motion of two athletes, it is possible to analyse the equivalent more simple motion, in the centre of mass reference system, of only one sham athlete gifted with a "reduced" body mass $\dfrac{m_1 m_2}{m_1 + m_2}$

b) In the centre of mass reference system, motion can be described by a two dimensional trajectory on the ground (mat) making use of the coordinates: r e θ.

C) Instead of solving the integral of motion by differential equations, it is better to use for the solution the Lagrangian of the system that is potential and kinetic energy.

To single out the general class among many potentials which will describe the common part of interaction, it is better to study the simplest kind of motion with constant angular momentum.

In this case the bi-dimensional trajectory can be treated as one-dimensional because

$$\theta = \dfrac{l}{mr^2}$$

and the interaction force F(r) will be function of distance between sham athlete and Centre of mass of Couple of Athletes system, that is of the sham potential $V' = V(r) + \dfrac{1}{mr^2}$ with

$$V(r) = -K r^{-a}$$

and the ά parameter a will take integral values 0,1,2,3, ...

The sham potential V'(r) will belong to one of subsequent classes of attractive potentials Fig.( 1 ).

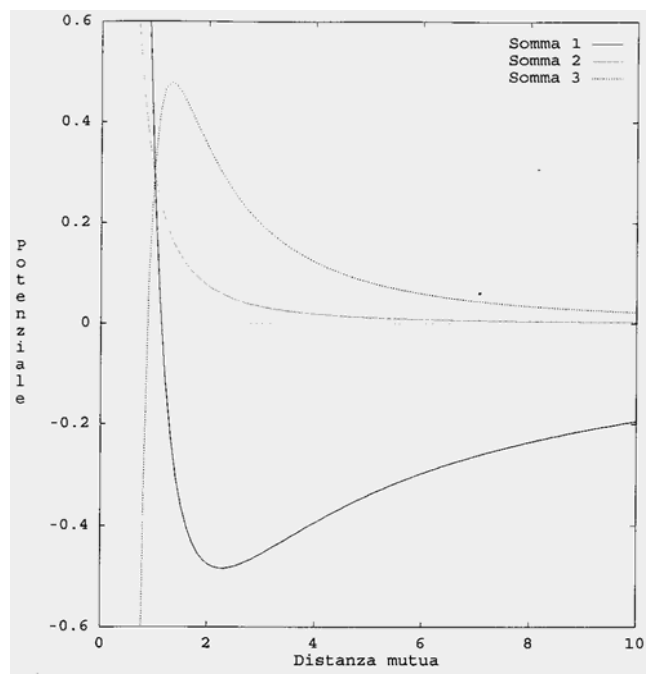

Fig 1  attractive sham potentials

This example shows very clearly that only attractive-repulsive potentials as $V'(r_1)$ will be useful to describe the common part of interaction during contest.

**VI POTENTIAL AND INTERACTION IN THE CENTER OF MASS REFERENCE SYSTEM**

The general potential which will describe the interaction will have the general exponential form:

$V' = r^{-a} + r^{-2a}$ From previous considerations, it is possible to declare that the common part of interaction can be described by the curves family showed in a generalised Morse's potential :

$V = D\left(e^{-2a(r-ro)} - 2e^{-a(r-ro)}\right)$ obviously V' is a particular expansion of this expression.

The specification of a general form of interaction potential, is able to give us a lot of useful information:

1) $r_o$ is the equilibrium distance (grip distance in wrestling ).
2) D is the mechanical potential energy in the equilibrium point equal to mechanical mean energy valued in terms of oxygen consumption as $\eta O_2$.
3) It is possible to evaluate the constant a expanding the potential near the minimum point. We get in this case the connection with the harmonic term of expansion $Da^2(r - r_o)^2 = E_c$ or

$$a = \frac{1}{L}\sqrt{\frac{E_c}{D}}$$

To know the potential let us go back to the Algebraic expression of force

$$F = ma = 2aD\left(e^{-ar} - e^{-2ar}\right)$$

To single out the common part of the interaction as a " two body problem in the central field" allows us to utilize an important result of classica1 physics about the mean time value of a few variables ( Virial's Theorem ) ; both for motion and interaction it guarantees that, if the generalized force F is a sum of friction and central forces , the mean kinetic energy of the system in time is independent from friction forces :

$$\overline{T} = -1/2 \frac{\partial V}{\partial r} r \approx \frac{\eta}{e}\overline{O_2} - \overline{V} = \eta_1 \overline{O_2} - \overline{V}$$

Where $\eta_1$ is the global efficiency of contest, always smaller than $\eta$.

The conservation of mean mechanical energy in time, on the basis of Virial's Theorem, allows us to obtain the expression of shifting velocity (2).

$$\dot{r} = \sqrt{\frac{2}{m}\left[\frac{\eta_1 O_2}{\tau}\frac{r^2\theta^2}{d} - D\left(e^{-ar} - 2e^{-2ar}\right)\right]}$$

The limit for r→ O of this expression allows us to calculate the attack velocity, at the instant of impact (r=O) which can be expressed (3) with regard to attacking oxygen consumption:

$$\lim_{r \to 0} \dot{r} = \sqrt{\frac{2D}{m}} = \sqrt{\frac{2\eta O_2}{m}}$$

**VII MOTION IN THE LABORATORY REFERENCE SYSTEM**

A) "Couple of Athletes" closed system.
B) "Couple of Athletes" open system.

***A) "Couple of Athletes" closed system.***
This system achieves "random" shifting by changing couple velocity direction in push/pull forces produced by Athletes themselves to generate specific "situation" in order to apply winning techniques.
In this case "random" means that statistically there is not a preferential shifting direction.
The motion can be accomplished by friction between soles and ground on the base of the III° principle of Dynamics; the general equation describing the situation is the II° Newton's Law  ma= F .
In the generalized force F will appear both friction and push/pull contribution.
The friction component is proportional to the velocity $F_a$ = -μv . The changing in velocity and direction produced by push/pulls are created by resultant of force developed by the two Athletes themselves.
They are, with regard to the whole contest time, acting impulses in very short intervals of time. Consequently the single variation can be described by Dirac's of the impulse u from the elementary force, where u means the mechanical momentum variation Δv m.
The resultant will be the algebraic sum of the push/pull forces (8), in which the random changes in direction will be evaluated as the variation ( ± 1) j of the elementary  force.
The whole force is : $\phi(t) = u \sum_j \delta(t - t_j)(\pm 1)_j = F'$

Then the generalized force is F = $F_a$ + F' and the general equation of the motion has the well-known structure of Langevin's Equation:

$$\dot{v} = -\frac{\mu}{m}v + \frac{u}{m}\sum_{j}(\pm 1)_j \delta(t-t_j) = F_a + F'$$

Because the push/pd resultant is "random" it is not possible to forecast the trajectory in only one contest, but the statistical analysis of many contests will be able to have information about the system behaviour.

1) Because the direction changes have the same probability, that is, over many contest there is not a preferred direction, then the mean value of F' in a random sequence of directions will be zero ( F' ) = 0

2) The mean over time and directions of two push/pull forces product give us information about force variation in time (8):

$$\langle F'(t)F'(t')\rangle = \frac{u^2}{m^2}\langle(\pm 1)_j(\pm 1)_j \delta(t-t_j)\delta(t'-t_j)\rangle = \frac{u^2}{m^2 t_o}\delta(t-t'_j)$$

Checking these conditions allows us to see easily that the motion of the Athletes can be described in terms of statistical mechanics as Brownian motion over an unlimited surface. The motion equation can be solved ( with constant variation methods ) The solution states that the system velocity is directly proportional to push/pull impulse u and inversely proportional to total mass m, then the biggest Athletes move themselves, statistically, at less velocity. $v(t) \propto \frac{u}{m}$

In these cases it is correct to evaluate only mean values of the quantities, for example the correlation function $\langle v(t)v(t')\rangle$ gives us the delay time of measurable velocity variation (8)

The solution is $\langle v(t)v(t')\rangle = \frac{m}{2\mu}C\left[e^{-\frac{\mu}{m}(t-t')} - e^{-\frac{\mu}{m}(t+t')}\right]$ if we think of steady state

(t+t')>> (t-t') the result is $\langle v(t)v(t')\rangle = \frac{m}{2\mu}Ce^{-\frac{\mu}{m}(t-t')}$ and the delay time is $t^* = \frac{m}{\mu}$ directly proportional to the Athletes' mass (8). If we put zero as the starting speed then it is possible to evaluate the kinetic energy mean value of more contests (8):

$$\frac{m}{2}\langle v^2 \rangle = \frac{m^2}{4\mu}C\left(1-e^{-\frac{2\mu}{m}t}\right)$$ for t→ ∞. This expression speedily tends to zero and then it is possible to write the steady state relationship $\frac{m}{2}\langle v^2 \rangle = \frac{m^2}{4\mu}C$

The C constant can be evaluated by a modified Einstein's method for the classic Brownian motion. Therefore if we think that the Athlete biosystem shows one of the lowest working efficiency or $\frac{L}{O_2} = \eta \ll 1$, then (8) it is possible to write $\frac{m}{2}\langle v^2 \rangle = \frac{m^2}{4\mu}C = \eta O_2$ or

$$C = \frac{4\mu}{m^2}\eta O_2$$

From this equation it is possible to get (8) the square momentum is directly proportional to friction and to overall oxygen consumption $u^2 = 4\mu t_0 \eta O_2$, while the correlation velocity function takes (8) the value $\langle v(t)v(t') \rangle = \frac{2}{m}\eta O_2 e^{-\frac{\mu}{m}(t-t')}$ from which the speedy fluctuation is inversely proportional to mass m , directly proportional to oxygen consumption and independent of friction.

### B) "Couple of Athletes" open system.

We have proved that the "bound" system, on an unlimited surface, moves with a Brownian motion; on the contest surface it is reasonable to think, for the overlap of effects, that the motion would be shaped by reflections from the surface edges (expressible as central field) plus the Brownian motion previously founded. Or alternatively the motion belongs to the class of "generalized Brownian motions" (2).
In the case of Couple of Athletes open system, the whole is given by two subsystems parted ( the Athletes ) which are in unstable equilibrium. If the biomechanical athlete is modelized as the classic model of inverted pendulum, then the oscillations will be the best way to struggle with gravity force. In this case, with this contrivance, the two biomechanical athletes will be regarded, with good approximation, in stable equilibrium; so to the whole Couple of Athletes open system can be extended the methodology first applied to the Couple of Athletes closed system ( with however, bigger mutual distance and with oscillations annulling the gravity force ).
Also in this condition the whole system moves, in the laboratory reference system, as generalized Brownian motion.

In the laboratory reference system, the equations of the motion of a single athlete considered as inverted pendulum have been taken from several authors ( Pedotti 1980, Mc Ghee 1980, Mc Mahon 1981 ).

Narrowing our study, for simplicity, to the bi-dimensional path on the ground and not to the tri-dimensional in space, and remembering the Brownian shifting (3) this relation will be valid :

$$r\theta = \sqrt{dt} \Rightarrow \frac{d}{\theta^2 r^2} = \frac{1}{t} \quad \text{or} \quad v = \frac{d}{4\pi^2 r^2}$$

It is possible to understand the physical meaning of this frequency in the limits r → ∞ and r→ 0, in this way we can deduce that it is the contact frequency ( or attack frequency ) or as function of measurable parameters:

$$v = \frac{1}{2\pi^2 r}\sqrt{\frac{2\eta 0_2}{m}} = \frac{v_i}{2\pi^2 r}$$

**VIII EXPERIMENTAL CHECK ( VERIFICATION- VALIDATION)**

The results about shifting trajectories study for Couple of Athletes system during contest, analysed by statistical mechanics techniques concerning "random" situations not repeatable in time, with a definite probability frequency, could appear only a theoretical exercise if they were not verified and validated with experimental results.

The next illustration shows "dromograms" from judo (1) championships which took place in Japan in 1971 and it is evident that there is not a preferential direction in push/pull forces over time. ( Fig. 2 )

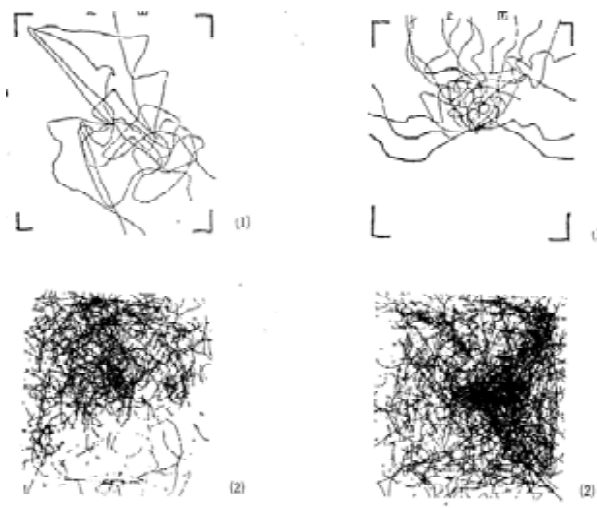

Fig 2  1,2, 7, 12 judo contest motion patterns in 1971 Japan championships.

## IX PHYSICAL PRINCIPLES AND INTERACTION TRAJECTORIES

"Couple of Athlete" closed system.

The interaction second face was solved by the author in the years 1985-1987 and led to the corollaries, about the use of forces in space, (static conditions), with the analysis of flight paths and symmetries ( dynamic conditions ) and with the identification of the basic physical principles of throwing techniques.

Using Galileo's principle of relativity it will be possible to extend the validity of the known results from static to dynamic conditions of contest.

1) Couple of forces techniques

For this class of techniques, we will apply the principle of simultaneous actions; then the problem of body motion in space can be simplified in the summation of a flat motion in sagittal or frontal plane, plus an eventually simplified motion in space.

So the first motion, produced by the application of the principal couple of forces, is a rotary flat motion independent of gravity force.

While the second ( applied in another group of the same class of techniques ) is the summation of motions produced by gravity force plus a secondary couple of forces, acting in a perpendicular plane to the gravitational field.

In terms of variational analysis the first motion flight path is obtained from the "extremals" of general function. $I = \int_{x1}^{x2} y^r \left(1 + y'^2\right)^{1/2} dx$ for r=-1 with solutions x=a- b sinθ e y = c – b cosθ in this case the "external" is the circle arc of radius b and centre B ( a,c).

While the second flight path is, with good approximation, the parabola arc with vertex V coincident with the rotation centre B of principal couple of forces Fig 3,4,5

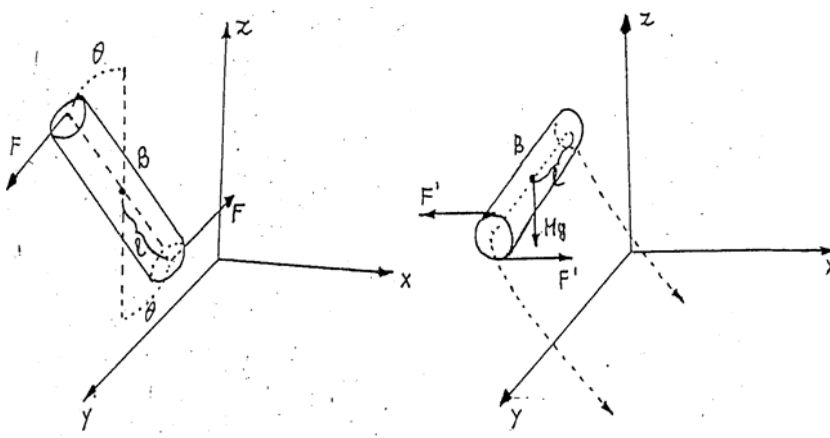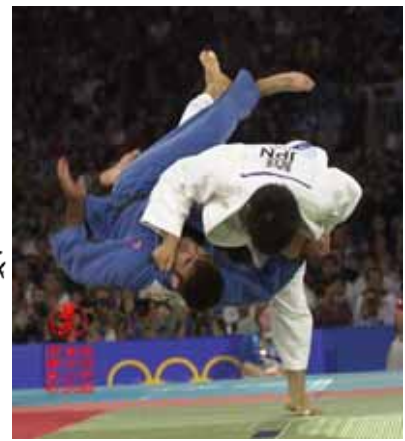

Fig 3,4,5 Technique of couple of forces –Uchi Mata

2) Techniques of physical lever.

Considering the biomechanical athlete as a rigid cylinder, applying the stopping point ( fulcrum), then the starting impulse must be regarded as necessary and sufficient to perform the unbalance, that is to shift the baricentral perpendicular out from the support base, while it sets to the adversary's body a rotational momentum under the condition

$$\frac{1}{2}I_z\omega^2_z \prec\prec 2Mgl$$

So the athlete can be assimilated to a symmetric heavy top, falling down in the gravitational field.

Because the starting impulse is acting during a short time lapse, the trajectory, in a force field, is given by the solution of the variational principle :

$$\delta\int_{q1}^{q2}\sum_j p_j dq_j - \delta\int_{t1}^{t2} H(p,q)dt$$

In our case the external field is conservative, then it is possible to apply the principle of minimum action, that is: 
$$\Delta\int_{t1}^{t2}\sum_j p_j q_j dt = \Delta\int_{t1}^{t2}\tau dt = \tau\Delta\int_{t1}^{t2} dt = \Delta(t_2 - t_1)$$

The body will go along the path of the least transit time. The Jacobi form of the principle of minimum action gives us other information:

$$\Delta\int_{t2}^{t2}\tau dt = \Delta\int_{\rho2}^{\rho2}\sqrt{H - V(\rho)}d\rho = 0$$

The p parameter measures the length of the path and makes sure the body will go along a geodetic of a special symmetry. In this case it is possible to show that it is going along a spiral arc , geodetic of a cylindrical symmetry.  Fig 6,7

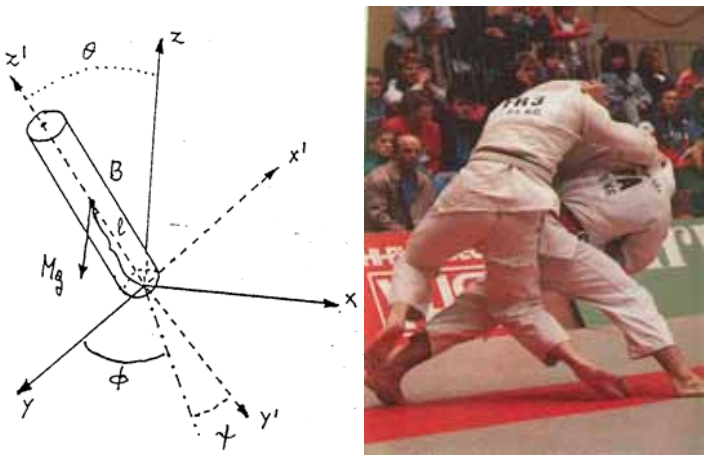
Fig 6,7 Technique of lever  Tai Otoshi

*X Probabilistic analysis of interaction*

"Couple of Athlete" open system.

In the study of the specific interaction for the Couple of Athletes open system, the attack mechanics guarantees that there are only two solutions: to be successful or not, faced to a defence which is effective or not.

Then comes the question of what probability of success have defence and attack, and how can they be connected whit each other by the probabilistic analysis.

In the first approximation direct attacks, that is without "shams" or "combinations" are independent. That means we have a series of Bernoulli tries.

By applying the binomial distribution to the attack, the probability of having 2 successes every 10 tries over the 12 possibilities of attacks is:

$$A = \frac{10!}{2!(10-2)!}\left(\frac{1}{12}\right)^2\left(1-\frac{1}{12}\right)^{10-2} = 0.155$$

While the defence probability to have 2 success every 10 tries over the 8 kind of defence is:

$$P = \frac{10!}{2!(10-2)!}\left(\frac{1}{8\pi}\right)^2\left(1-\frac{1}{8\pi}\right)^{10-2} = 0.05$$

The mathematical probability of 2 successes every 10 direct attacks is 16% while 2 every 10 defence is 5%.

So this case the probabilistic analysis shows us that in Couple of Athletes open system, attack will be easier and better than defence.

**XI CONCLUSIONS**

The biomechanical analysis of contest sports competitions has given us some very important results:

a) the motion of Couple of Athletes system is a Brownian generalized motion;

b) interaction between athletes can be subdivided into two steps: the first is common to all contest sports (shortening of mutual distance); the second step is peculiar to each sport.

Ex. long distance sports: direct blows to sensible or conventional body points; very short distance sport: tools for throwing the adversaries by two physical principles;

c) the kinetic energy of the athlete depends on oxygen consumption and the athlete's efficiency;

d) The capability of changing velocity does not depend on friction; it is inversely proportional to the mass and directly dependent on oxygen consumption and efficiency;

e) the variation time of velocity is dependent on the mass and inversely proportional to friction;

f) the attack velocity at contact is given by square root of double of oxygen consumption multiplied by efficiency divided by body or limb mass;

g) the measure of variation of push/pull force is related to the measure of the friction in direct proportion to oxygen consumption;

h) the flight path of thrown athlete is a geodetic of three specified symmetry or their linear composition;

i) the couples of forces techniques are independent of friction; it is possible to use them whatever the shifting velocity is;

j) the physical lever techniques depend on friction ; that means it is possible to use them only for stopping the adversary;

k) the techniques of couple of forces are energetically the best;

l) among the techniques of physical lever, the maximum a m are energetically the best;

m) attack frequency is directly proportional to impact velocity and indirectly to mutual distance;

n) attack frequency is directly proportional to kinetic energy for time and inversely to mass and square of distance;

o) the direct attack blow has a success probability of 66% compared to defence .